\documentclass[12pt]{article}

\usepackage{graphics}

\usepackage{amssymb}

\newcommand{\QED}{\mbox{\rule[-1.5pt]{6pt}{10pt}}}

\newcommand{\Ran}{\mathrm{Ran}\,}

\newcommand{\R}{\mathbb{R}}

\newcommand{\HH}{{\mathcal H}}

\newtheorem{claim}{Claim}[section]

\newtheorem{theorem}[claim]{Theorem}

\newtheorem{corollary}[claim]{Corollary}

\newtheorem{proposition}[claim]{Proposition}

\newtheorem{lemma}[claim]{Lemma}

\newtheorem{remark}[claim]{Remark}

\begin{document}

\title{A Product Formula Related to \\ Quantum Zeno Dynamics}

\author{Pavel Exner$^{a,b}$ and Takashi Ichinose$^{c}$}

\date{}

\maketitle

\begin{quote}

{\small \em a) Department of Theoretical Physics, Nuclear Physics
Institute, \\ \phantom{e)x}Academy of Sciences, 25068 \v Re\v z,
Czech Republic \\
 b) Doppler Institute, Czech Technical University, B\v{r}ehov{\'a}
7,\\
\phantom{e)x}11519 Prague, Czech Republic \\
 c) Department of Mathematics, Faculty of Science, Kanazawa \\
\phantom{e)x}University, Kanazawa 920-1192, Japan \\
 \rm \phantom{e)x}exner@ujf.cas.cz, ichinose@kenroku.kanazawa-u.ac.jp}
\vspace{8mm}

\medskip\noindent
{\small We prove a product formula which involves the unitary
group generated by a semibounded self-adjoint operator and an
orthogonal projection $P$ on a separable Hilbert space $\HH$, with
the convergence in $L^2_\mathrm{loc}(\R;\HH)$. It gives a partial
answer to the question about existence of the limit which
describes quantum Zeno dynamics in the subspace
\hbox{$\mathrm{Ran}\,P$}. The convergence in $\HH$ is demonstrated
in the case of a finite-dimensional $P$. The main result is
illustrated in the example where the projection corresponds to a
domain in $\R^d$ and the unitary group is the free Schr\"odinger
evolution.}
\end{quote}


\section{Introduction}

The fact that the decay of an unstable system can be slowed down,
or even fully stopped in the ideal case, by frequently repeated
measurements checking whether the system is still undecayed was
noticed first by Beskow and Nilsson \cite{BN}. It was only decade
later, however, when Misra and Sudarshan \cite{MS} caught the
imagination of the community by linking the effect to the
well-known Zeno aporia about a flying arrow. While at first the
subject was rather academical, in recent years the possibility of
observing Zeno-type effects experimentally has become real and at
present there are scores of physical papers discussing this topic.

On the mathematical side, the first discussion of the continuous
observation appeared in \cite{Fr}. Two important questions,
however, namely the existence of Zeno dynamics and the form of its
effective Hamiltonian have been left open both in this paper and
later in \cite{MS}. The second problem is particularly important
when the subspace into which the state of the system is repeatedly
reduced has dimension larger than one. A partial answer was given
in \cite[Sec.~2.4]{Ex} where it was shown that the results of
Chernoff \cite{Ch1, Ch2} allow to determine the generator of the
Zeno time evolution naturally through the appropriate quadratic
form.

Our interest to the problem was rekindled by a recent paper by
Facchi et al. \cite{FP} who studied the important special case
when the presence of a particle in a domain of $\Omega
\subset\R^d$ is repeatedly ascertained. Using the method of
stationary phase the authors showed that the Zeno dynamics
describes in this case the free particle confined to $\Omega$,
with the hard-wall (Dirichlet) condition at the boundary of the
domain. The result cannot be regarded as fully rigorous, because
detailed properties of the convergence are not worked out, but the
idea is sound without any doubt.

In the present paper we combine the results of \cite{Ch1, Ch2}
with that of Kato \cite{Ka2} to address this question in a general
setting. We show that if the natural effective Hamiltonian
mentioned above is densely defined --- which is a nontrivial
assumption --- then the Zeno dynamics exists and the said operator
is its generator in a topology which includes an averaging over
the time variable -- cf.~Theorem~\ref{main} for exact statement (a
part of the present result given in Corollary~\ref{cor2} was
announced in \cite{EI}). Our conclusion cannot be thus regarded as
fully satisfactory from the mathematical point of view, because
the natural topology to be used here is given by the norm of the
Hilbert space, and in this respect an important part of the
problem remains open. We demonstrate, however, the strong
convergence in $\HH$ for the particular case when the projections
involved are finite-dimensional -- cf.~Theorem~\ref{fin}. On the
other hand, from the physical point of view the result given in
Theorem~\ref{main} is quite plausible taking into account that any
real measurement is burdened with errors -- see
Remark~\ref{phys-conv} below.

We will formulate the theorems together with their corollaries in
the next section. Theorem~\ref{main} will be then proven in
Sections \ref{s-proof} and \ref{s-proof-of-L1}, Theorem~\ref{fin}
in Section~\ref{finite}. As an example we discuss in the
concluding section reduction of a free dynamics to a domain in
$\R^d$ by permanent observation. We will establish that the Zeno
generator mentioned above is in this case the Dirichlet Laplacian,
obtaining thus in a different way the result of the paper
\cite{FP}.


\setcounter{equation}{0}

\section{The main result} \label{s-main}

Throughout the paper $H$ will be a nonnegative self-adjoint
operator in a {\it separable} Hilbert space $\HH$, and $P$ will be
an orthogonal projection. The nonnegativity assumption is made for
convenience; our main result extends easily to any self-adjoint
operator $H$ bounded from below as well as one bounded from above,
i.e. to each semi-bounded self-adjoint operator in $\HH$.

Consider the quadratic form $u \mapsto \|H^{1/2}Pu\|^2$ with form
domain $D[H^{1/2}P]$. Note that $H^{1/2}P$ involved here is a
closed operator and $HP$ has the same property. Let $H_P :=
(H^{1/2}P)^*(H^{1/2}P)$ be the self-adjoint operator associated
with this quadratic form. In general, $H_P$ may not be densely
defined in which case it is a self-adjoint operator in a closed
subspace of $\HH$. More specifically, it is obviously defined and
acts nontrivially in a closed subspace $\Ran P$ of the closure of
the form domain $D[H^{1/2}P]$, while in the orthogonal complement
$(\Ran P)^\perp$ it acts as zero.

The quadratic form $u \mapsto \|H^{1/2}Pu\|^2$ defined on
$D[H^{1/2}P]$ is a closed extension of the form $u \mapsto \langle
Pu,HPu \rangle$ defined on $D[HP]$, but the former is not in
general the closure of the latter. Indeed,  if $H$ is unbounded,
$D[H]$ is a proper subspace of $D[H^{1/2}]$. Take $u_0 \in D
[H^{1/2}]\backslash D[H]$ such that the vector $H^{1/2}u_0$ is
nonzero, and set $P$ to be the orthogonal projection onto the
one-dimensional subspace spanned by $u_0$. Taking into account
that $D[HP] = \{u \in {\cal H};\,\, Pu \in D[H]\}$ which $u_0
=Pu_0$ does not belong to, we find $HPu = 0$ for $u \in D[HP]$,
while $H^{1/2}Pu_0 = H^{1/2}u_0 \not= 0$ by assumption.

To describe our results, we denote by
$L^2_\mathrm{loc}([0,\infty); {\cal H}) =
L^2_\mathrm{loc}([0,\infty)) \otimes {\cal H}$ the Fr\'echet space
of the ${\cal H}$-valued strongly measurable functions $v(\cdot)$
on $[0,\infty)$ such that $\|v(\cdot)\|$ is locally square
integrable there, with the topology induced by the semi-norms
$v\mapsto \bigl(\int_0^{T_{\ell}} \|v(t)\|^2 dt\bigr)^{1/2}$ for a
countable set $\{T_{\ell}\}_{\ell =1}^{\infty}$ of increasing
positive numbers accumulating at infinity, $\lim_{\ell \rightarrow
\infty} T_{\ell} = \infty$. In a similar way one defines the
Fr\'echet space $L^2_\mathrm{loc}(\R; {\cal H}) =
L^2_\mathrm{loc}(\R) \otimes {\cal H}$.

\medskip
Our main result can be stated as follows:

 \begin{theorem} \label{main}
Let $H$ be a nonnegative self-adjoint operator on a separable
Hilbert space $\HH$ and $P$ an orthogonal projection. Let $t
\mapsto P(t)$ be a strongly continuous function whose values are
orthogonal projections in ${\cal H}$, defined in some neighborhood
of zero, with $P(0) =: P$. Moreover, suppose that $D[H^{1/2}P(t)]
\supset D[H^{1/2}P]$ and $\lim_{t \rightarrow 0} \|H^{1/2}P(t)v\|
= \|H^{1/2}Pv\|$ holds for $v \in D[H^{1/2}P]$. If the operator
$H_P$ specified above is densely defined in the whole Hilbert
space $\HH$, then for every $f \in {\cal H}$ and $\varepsilon=\pm
1$ it holds that
 \begin{eqnarray}
&&[P(1/n)\exp(-i\varepsilon tH/n) P(1/n)]^n f
\longrightarrow
\exp(-i\varepsilon tH_P)\,Pf\,, \label{product-s} \\
&&[P(1/n)\,\exp(-i\varepsilon tH/n) ]^n f
\longrightarrow
\exp(-i\varepsilon tH_P)\,Pf\,, \label{product-ns}\\
&&[\,\exp(-i\varepsilon tH/n)\,P(1/n)]^n f
\longrightarrow
\exp(-i\varepsilon tH_P)\,Pf\,, \label{product-ns2}
 \end{eqnarray}
in the topology of $L_\mathrm{loc}^2(\R; {\cal H})$ as
$\:n\rightarrow \infty$.
 \end{theorem}

\noindent Note that $H_P$ differs in general from the operator
$PHP$, which may not be self-adjoint in $\HH$, nor even closed,
because $PH$ is not necessarily closed, though $HP$ is. $H_P$ is a
self-adjoint extension of $PHP$. The requirement of the theorem
that $H_P$ is densely defined in $\HH$ means nothing else but
that the domain $D[H^{1/2}P]$ of the quadratic form in question is
dense in $\HH$.

Note also that for $\varepsilon =1$, the theorem concerns a
nonnegative self-adjoint operator $\varepsilon H = H$, while  for
$\varepsilon =-1$, we get  product formulae for the non-positive
self-adjoint operator $\varepsilon H = -H$. Moreover, the result
is preserved when $H$ is replaced with a shifted operator $H+cI$,
i.e. for any semi-bounded self-adjoint operator in a separable
Hilbert space.

An important particular case, most often met in the applications,
concerns the situation when the projection-valued function is
constant.
 \begin{corollary} \label{cor1}
Let $H$ be a self-adjoint operator bounded from below in a
separable Hilbert space $\HH$ and $P$ an orthogonal projection. If
the operator $H_P$ specified above is densely defined, then for
every $f \in {\cal H}$ and $\varepsilon = \pm 1$ we have in the
topology of $L_\mathrm{loc}^2(\R; {\cal H})$ the limiting relation
 \begin{equation}
[P\exp(-i\varepsilon tH/n) P]^n f \longrightarrow
  \exp(-i\varepsilon tH_P)\,Pf \label{c:product-s}
\end{equation}
for $\:n\rightarrow \infty$ as well as its nonsymmetric
counterparts obtained by setting $P(1/n)=P$ in (\ref{product-ns})
and (\ref{product-ns2}).
 \end{corollary}

>From the viewpoint of quantum Zeno effect described in the
introduction the optimal result would be a strong convergence on
$\mathcal{H}$ for a fixed value of the time variable, moreover
uniformly on each compact interval in $t$. Our Theorem~\ref{main}
implies the following weaker result on pointwise convergence.

 \begin{corollary} \label{cor2}
Under the same hypotheses as in Theorem~\ref{main}, there exist a
set $M \subset \R$ of Lebesgue measure zero and a strictly
increasing sequence $\{n'\}$ of positive integers along which we
have
 \begin{eqnarray}
&& [P(1/n')\exp(-i\varepsilon tH/n') P(1/n')]^{n'} f
 \longrightarrow
 \exp(-i\varepsilon tH_P)\,Pf\,, \label{ae:product-s} \\
&& [P(1/n')\,\exp(-i\varepsilon tH/n') ]^{n'} f
 \longrightarrow
\exp(-i\varepsilon tH_P)\,Pf\,, \label{ae:product-ns}\\
&& [\,\exp(-i\varepsilon tH/n')\,P(1/n')]^{n'} f
 \longrightarrow
\exp(-i\varepsilon tH_P)\,Pf\,, \label{ae:product-ns2}
 \end{eqnarray}
for every $f \in \mathcal{H}$, strongly in ${\cal H}$ for all $t
\in \R\setminus M$.
 \end{corollary}

As we have indicated above, one need not resort to subsequences in
the particular case when the projections involved are
finite-dimensional.

\begin{theorem} \label{fin}
In addition to the hypotheses of Theorem~\ref{main}, assume that
the orthogonal projection $P$ as well as $P(t)$ is of finite
dimension. Then \\ [.2em]
(i) the formulae (\ref{product-s})--(\ref{product-ns2}) hold in
the norm of ${\cal H}$ as $n\rightarrow \infty$, uniformly on each
compact interval of the variable $t$ in $\R\setminus \{0\}$, \\
[.1em]
(ii) it also holds for $\varepsilon = \pm 1$ that as $n\rightarrow
\infty$,
 \begin{eqnarray*}
&& [P(t/n)\exp(-i\varepsilon tH/n) P(t/n)]^{n}
 \longrightarrow
 \exp(-i\varepsilon tH_P)\,P\,, \label{} \\
&& [P(t/n)\,\exp(-i\varepsilon tH/n)]^{n}
 \longrightarrow
\exp(-i\varepsilon tH_P)\,P\,, \label{}\\
&& [\,\exp(-i\varepsilon tH/n)\,P(t/n)]^{n}
 \longrightarrow
\exp(-i\varepsilon tH_P)\,P\,, \label{}
 \end{eqnarray*}
strongly on ${\cal H}$, uniformly on each compact interval in the
variable $t\in\R$.
\end{theorem}

\medskip
 Before proving Theorems~\ref{main} and \ref{fin} and
 Corollary~\ref{cor2} let us comment briefly on several other
 aspects of the result.

 \begin{remark} \label{phys-conv}
{\rm While the necessity to pick a subsequence makes the pointwise
convergence result weaker than desired, let us notice that from
the physical point of view the convergence in
$L_\mathrm{loc}^2(\R; {\cal H})$ can be regarded as satisfactory.
The point is that any actual measurement, in particular that of
time, is burdened with errors. Suppose thus we perform the Zeno
experiment on numerous copies of the system. The time value in the
results will be characterized by a probability distribution
$\phi:\:\R_+ \to\R_+$, which is typically a bounded, compactly
supported function -- in a precisely posed experiment it is
sharply peaked, of course. Corollary~\ref{cor1} then gives
 \begin{equation}
\int \phi(t)\, \left\|\,[P\exp(-i\varepsilon tH/n) P]^n f -
  \exp(-i\varepsilon tH_P)\,P f \right\|^2\,dt \to 0 \label{physZeno}
\end{equation}
as $n\to\infty$, in other words, the Zeno dynamics limit is valid
after averaging over experimental errors, however small they are.}
 \end{remark}

 \begin{remark}
{\rm While the proof of strong convergence in ${\cal H}$ in
Theorem~\ref{main} and Corollaries~\ref{cor1} and \ref{cor2}
remains elusive without the finite-dimension assumption, such a
claim can be easily established in the orthogonal complement of
the subspace $P\HH$. Indeed, taking $f\in Q{\cal H}$, where
$Q:=I-P$, we have
\begin{eqnarray*}
 (P(1/n)e^{-i\varepsilon tH/n} P(1/n))^n f
  &\!=\!& (P(1/n)e^{-i\varepsilon tH/n} P(1/n))^n P(1/n)Qf\,,\\
 (e^{-i\varepsilon tH/n} P(1/n))^n f
 &\!=\!& (e^{-i\varepsilon tH/n} P(1/n))^n P(1/n)Qf\,,
\end{eqnarray*}
which converge to zero, uniformly on each compact $t$-interval in
$\R$, as $n\rightarrow \infty$, because $P(\tau)
\stackrel{\scriptsize{s}}{\rightarrow} P$ as $\tau \rightarrow 0$.
This gives the result for (\ref{ae:product-s}) and
(\ref{ae:product-ns2}), while for (\ref{ae:product-ns}) one has to
employ in addition the relation (\ref{diff:ns-s}) below. }
 \end{remark}

 \begin{remark}
{\rm The fact that the product formulae require $H_P$ to be
densely defined is nontrivial. Recall the example of
\cite[Rem.~2.4.9]{Ex} in which $H$ is the multiplication operator,
$(H\psi)(x)=x\psi(x)$ on $L^2(\R_+)$, and $P$ is the
one-dimensional projection onto the subspace spanned by the vector
$\psi_0:\: \psi_0(x)= [(\pi/2)(1\!+\!x^2)]^{-1/2}$. In this case
obviously $H_P$ is the zero operator on the domain $D[H_P]=
\{\psi_0\}^\perp$. On the other hand, $P\, e^{-itH}P$ acts on
$\Ran P$ as multiplication by the function
 $$ v(t):= e^{-t} -{i\over\pi} \left\lbrack e^{-t}\,
 \overline\mathrm{Ei}(t) - e^{t}\,\mathrm{Ei}(-t) \right\rbrack =
 1+{2i\over\pi}\, t\ln t + \mathcal{O}(t), $$
where $E_{i}(-t)$ and $\overline{E}_i(t)$ are exponential
integrals \cite{AS}; due to the rapid oscillations of the
imaginary part as $t \downarrow 0$ a pointwise limit of $v(t/n)^n$
for $n\to\infty$ does not exist. Notice also that different limits
may be obtained in this example along suitably chosen subsequences
$\{n'\}.$ }
 \end{remark}

 \begin{remark}
{\rm In their recent study of Trotter-type formulae involving
projections Matolcsi and Shvidkoy \cite{MaS} presented two
examples in which expressions of the type $[\exp(-iH/n) P]^n$ do
not converge strongly. This result does not answer the question,
however, whether the product expressions considered here converge
in the strong topology of $\mathcal{H}$ or not, because our
assumptions are not satisfied there. In the first example of
\cite{MaS} the analogue of the operator $H_P$ is not densely
defined, in the second one $H$ is not semi-bounded. }
 \end{remark}


\setcounter{equation}{0}

\section{Proof of Theorem~\ref{main}} \label{s-proof}

We present the argument for $\varepsilon =1$, the case
$\varepsilon =-1$ can be treated similarly. We first prove
(\ref{product-s}) in (a), and next (\ref{product-ns}),
(\ref{product-ns2}) in (b).

\medskip
\noindent (a) Let us begin with the symmetric product case and
prove the formula (\ref{product-s}) with $\varepsilon =1$. We will
check the convergence in (\ref{product-s}) on an arbitrary compact
$t$-interval in the closed right half-line $[0,\infty)$. The proof
for $t$-intervals in the closed left half-line $(-\infty, 0]$ is
analogous, and in addition, it can be included in the case
$\varepsilon = -1$ with the convergence in (\ref{product-s}) on
compact $t$-intervals of the closed right half-line $[0,\infty)$.

Put $Q(t) := I-P(t)$ and $Q:= Q(0) = I- P(0) = I-P$, where $I$ is
the identity operator on ${\cal H}$. Since $H$ is nonnegative by
assumption, there exists a spectral measure $E(d\lambda)$ on the
nonnegative real line such that $H = \int _{0-}^{\infty} \lambda\,
E(d\lambda)$. For $\zeta \in {\mathbb{C}}$ with $\hbox{\rm
Re}\,\zeta \geq 0$ and $\tau > 0$, we put
\begin{equation} \label{red-evol}
 F(\zeta, \tau) = P(\tau)\,e^{-\zeta\tau H}P(\tau)\,,
\end{equation}
which is a contraction, and
\begin{equation} \label{red-evol-der}
 S(\zeta,\tau) = \tau^{-1}[I - F(\zeta,\tau)] =
 \tau^{-1}[I- P(\tau)\,e^{-\zeta\tau H}P(\tau)],
\end{equation}
which exists as a bounded operator on ${\cal H}$ with $\hbox{\rm
Re}\, \langle f,S(\zeta,\tau)f\rangle \geq 0$ for every $f \in
{\cal H}$. For definiteness we use here and in the following the
physicist convention about the inner product supposing that it is
antilinear in the first argument. For a non-zero $\zeta \in
{\mathbb{C}}$ with $\hbox{\rm Re}\, \zeta \geq 0$, we put also
 \begin{equation} \label{Htau}
 H(\zeta) := \zeta^{-1}[I- e^{-\zeta H}]\,.
 \end{equation}

\medskip
Each element $v(\cdot)$ in $L^2_\mathrm{loc}([0,\infty); {\cal
H})$ is an equivalence class such that any two representatives of
it are equal a.e. on $[0, \infty)$. However, at some places we
will not avoid an abuse of notation using for a particular
representative of such an element the same symbol $v(\cdot)$. At
the same time, in the following the convergence of a family of
vectors $v(\cdot,\tau)$ to $v(\cdot)$ in the topology of the space
$L^2_\mathrm{loc}([0,\infty); {\cal H}) =
L^2_\mathrm{loc}([0,\infty))\otimes {\cal H}$ as $\tau$ tends to
zero will be often written as $v(t,\tau) \longrightarrow v(t)$;
this will be the case when writing $v(\cdot,\tau) \longrightarrow
v(\cdot)$ would require to introduce a separate symbol for this
$v(t,\tau)$ the meaning of which is clear from the context.

The key ingredient of the proof is the following lemma.

 \begin{lemma} \label{L1}
$\:(I + S(it ,\tau))^{-1}$ converges to $(I + itH_P)^{-1}P$ as
$\tau \to 0$ strongly in $L^2_\mathrm{loc}([0,\infty);\, {\cal
H})$, in other words, for all $f \in {\cal H}$ and every finite
$T>0$ we have
 \begin{equation} \label{inv-gen}
 \int_0^T \|(I + S(it ,\tau))^{-1}f - (I + itH_P)^{-1}Pf\|^2\, dt
  \rightarrow 0\,, \quad \tau \to 0\,.
 \end{equation}
 \end{lemma}

\medskip

\noindent We postpone the proof of Lemma~\ref{L1} to the next
section. For the moment we will accept its claim and use it to
show that it implies the symmetric case (\ref{product-s}) of the
product formula in Theorem~\ref{main}.

\bigskip

To this end, let $\{m_n\}$ be a strictly increasing sequence of
positive integers, i.e. a subsequence of the sequence of all
positive integers. We have only to show that there exists a
subsequence $\{n'\}$ in any such sequence $\{m_n\}$ along which
(\ref{product-s}) holds. Then by a standard argument we can
conclude that (\ref{product-s}) actually holds along the sequence
of all positive integers $n$. For if this were not the case, there
would exist a subsequence $\{n'\}$ of strictly increasing positive
integers along which (\ref{product-s}) does not converge. However,
we see that there is a subsequence $\{n''\}$ of $\{n'\}$ along
which the convergence takes place to the same limit, which is a
contradiction.

Fix $\{m_n\}$ and $f \in {\cal H}$. Lemma~\ref{L1} holds, in
particular, along the sequence $\{\tau_n\}$ with $\tau_n :=
1/m_n$, and since $L^2$ convergence implies pointwise convergence
a.e. along a subsequence, there exist a subset $M_f$ of Lebesgue
measure zero of the variable $t$ in $[0,\infty)$ and a subsequence
$\{\tau_{f,n}\}$ of $\{\tau_n\}$, both dependent on $f$, such that
 $$
 (I+ S(it,\tau_{f,n}))^{-1}f \longrightarrow
          (I+ itH_P)^{-1}Pf
 $$
holds strongly in ${\cal H}$ for $t \in [0,\infty) \setminus M_f$.
Since ${\cal H}$ is separable by assumption, we can choose a
countable dense subset ${\cal D} = \{f_{\ell}\}_{\ell
=1}^{\infty}$ in ${\cal H}$. Then we infer that for $f_1 \in {\cal
D}$ there exist a set $M_1 := M_{f_1}$ of Lebesgue measure zero
and a subsequence $\{\tau_{1,n}\}$ of $\{\tau_n\}$ along which
$(I+ S(it,\tau_{1,n}))^{-1}f$ converges to $(I+ itH_P)^{-1}Pf$ for
every $t \notin M_1$. Next, for $f_2 \in {\cal D}$ there exist a
set $M_2 := M_{f_2}$ of Lebesgue measure zero and a subsequence
$\{\tau_{2,n}\}$ of $\{\tau_{1,n}\}$ along which $(I+
S(it,\tau_{2,n}))^{-1}f$ converges to $(I+ itH_P)^{-1}Pf$ for
every $t \notin M_2$. Proceeding in this way, we associate in the
$\ell$-th step with $f_{\ell} \in {\cal D}$ a set $M_{\ell} :=
M_{f_{\ell}}$ of Lebesgue measure zero and a subsequence
$\{\tau_{\ell,n}\}$ of $\{\tau_{\ell -1,n}\}$ along which $(I+
S(it,\tau_{\ell,n}))^{-1}f$ converges to $(I+ itH_P)^{-1}Pf$ for
every $t \notin M_{\ell}$.

Now we put $\tau'_n := \tau_{n,n}$ and  $n' := 1/\tau'_n$, so that
$\{n'\}$ is a subsequence of the strictly increasing sequence
$\{m_n\}$ of positive integers from which we have started. Then it
follows that for every $t \in [0,\infty) \setminus \cup_{\ell
=1}^{\infty} M_{\ell}\,$, the sequence $\{(I+
S(it,\tau'_{n}))^{-1}f\}$ converges to $(I+ itH_P)^{-1}Pf$
strongly in ${\cal H}$ as $\tau'_n \rightarrow 0$ for every $f \in
{\cal D}$, and therefore also in ${\cal H}$, because both $(I+
S(it,\tau_{\ell,n}))^{-1}$ and $(I+ itH_P)^{-1}P$ are bounded
operators on ${\cal H}$ with the norms not exceeding one. We
denote $M := \cup_{\ell =1}^{\infty} M_{\ell}$, which is, of
course, again a set of Lebesgue measure zero. In this way we have
found a subsequence $\{\tau'_n\}$ of $\{\tau_n = 1/m_n\}$ and an
exceptional subset $M$ of $[0,\infty)$ such that
\begin{equation} \label{resolconv}
 (I+ S(it,\tau'_{n}))^{-1}f = (I+ S(it,1/n'))^{-1}f
    \longrightarrow (I+ itH_P)^{-1}Pf
\end{equation}
strongly in ${\cal H}$ as $\tau'_n \rightarrow 0$ or $n'
\rightarrow \infty$ for every $f \in {\cal H}$ and for each fixed
$t \notin M$; it is important that $M$ is independent of $f$.

 \begin{lemma} \label{L2}
For the sequence $\{n'\}$ specified above and every $f \in \HH$ we
have
\begin{equation} \label{prodconv}
 [P(1/n')\exp(-itH/n') P(1/n')]^{n'}f
 \longrightarrow  e^{-itH_P}Pf
\end{equation}
as $\:n' \to\infty$ strongly in $\HH$ provided $\:t\notin M$.
\end{lemma}

\noindent Notice that this claim is in fact the ``symmetric'' part
of Corollary~\ref{cor2}.

\medskip\noindent
{\it Proof of Lemma~\ref{L2}:} We use arguments analogous to those
employed in derivation of Chernoff's theorem -- see \cite[Theorem
1.1]{Ch2}, \cite{Ch1} and \cite[Thm~IX.3.6]{Ka1}. We divide the
proof into two steps referring to $f$ belonging to $P\HH$ and to
its orthogonal complement.

Suppose first that $f \in P{\cal H}$. For $t \notin M$ and $\tau$
fixed, the operator $S(it, \tau)$ generates a strongly continuous
semigroup $\{\,e^{-\theta S(it, \tau)}:\: \theta \geq 0\,\} $ on
$\HH$, and the resolvent convergence (\ref{resolconv}) implies the
convergence of the corresponding semigroups
\cite[Thm~IX.2.16]{Ka1}, so we have
$$
 e^{-\theta S(it, 1/n')}f \stackrel{\scriptsize{s}}{\longrightarrow}
 e^{-i\theta t H_P}f
$$
in $\HH$ as $\:n'\to\infty$ for $\:t \notin M$, uniformly on each
compact interval of the variable $\theta \geq 0$. In particular,
choosing $\theta = 1$ we get for each $t \in [0,\infty)\setminus
M$
\begin{equation} \label{L3.3-semigrconv}
 e^{-S(it, 1/n')}f \stackrel{\scriptsize{s}}{\longrightarrow}
 e^{-it H_P}f\,, \quad n'\rightarrow \infty\,.
\end{equation}
The same equivalence implies for any $\lambda \geq 0$ and $t \in
[0,\infty)\setminus M$ that
$$
 (I+ \lambda S(it, 1/n'))^{-1}f
 \stackrel{\scriptsize{s}}{\longrightarrow}
 (I+i\lambda tH_P)^{-1}Pf\,,
$$
in particular, using the diagonal trick we obtain
\begin{equation} \label{L3.3-resolvconv}
 \Big(I+ \frac{1}{\sqrt{n'}} S(it, 1/n')\Big)^{-1}f \stackrel{\scriptsize{s}}
 {\longrightarrow}Pf \quad  \mathrm{as}\;\; n'\rightarrow \infty,
\end{equation}
for every $t \in [0,\infty)\setminus M$. Next we use
\cite[Lemma~2]{Ch1} which gives for any $g \in {\cal H}$ the
inequality
$$
\left\|F(it, 1/n')^{n'} g - e^{-n'(I-F(it,1/n'))}g\right\|
 \leq \sqrt{n'}\left\|(I-F(it, 1/n'))g\right\|\,.
$$
Choosing $g = \left(I+ \frac{1}{\sqrt{n'}} S(it,
1/n')\right)^{-1}\!f\:$ we infer that
\begin{eqnarray*}
&&\left\|\left[ F(it, 1/n')^{n'} - e^{-S(it,1/n')}\right]
       \left(1+ \frac{1}{\sqrt{n'}} S(it, 1/n')\right)^{-1}f
       \right\| \\
&& \quad \leq
 \left\|\left(I+ \frac{1}{\sqrt{n'}} S(it, 1/n')\right)^{-1}f
 - f\right\|\,,
\end{eqnarray*}
where the right-hand side tends to zero as $n' \rightarrow \infty$ by
(\ref{L3.3-resolvconv}). Using (\ref{L3.3-resolvconv}) once again
we get
\begin{equation} \label{L3.3-semigrconv2}
\left\|F(it, 1/n')^{n'} f - e^{-S(it,1/n')}f \right\|
 \longrightarrow  0\,.
\end{equation}
The sought relation (\ref{prodconv}) immediately follows from
(\ref{L3.3-semigrconv}) and (\ref{L3.3-semigrconv2}), since by
(\ref{red-evol}) we have $F(it,1/n')^{n'} = [P(1/n')\exp(-itH/n')
P(1/n')]^{n'}$.

The case $f \in Q\HH$ is easier being independent of the arguments
preceding Lemma~\ref{L2}. We have, along the sequence of {\it all}
positive integers $n$,
 $$ [P(1/n)\exp(-itH/n) P(1/n)]^n f \rightarrow 0
 $$
strongly in $\HH$ and for each $t\in[0, \infty)$, since $P(1/n)f =
P(1/n)Qf$ converges by assumption to $PQf =0$ as $n\to\infty$,
while $\exp(-itH_P)Pf = 0$. \QED

\medskip

This yields the sought result because $\{[P(1/n')\exp(-itH/n')
P(1/n')]^{n'}\}$ is a bounded sequence for any $t\ge 0$ and by
Lebesgue dominated-convergence theorem it tends to the expected
limit in $L_\mathrm{loc}^2([0,\infty); {\cal H})$. Using the
standard ``subsequence'' trick mentioned above we have thus shown
that Lemma~\ref{L1} implies the symmetric product formula
(\ref{product-s}) of Theorem~\ref{main}.

\bigskip

\noindent(b) Let us turn to the non-symmetric product-formula
cases, i.e. to prove that (\ref{product-s}) implies
(\ref{product-ns}) and (\ref{product-ns2}).

\medskip\noindent
{\it Proof of (\ref{product-ns}):} We employ the standard
notation, $[U,S] = US - SU$, for the commutator of bounded
operators $U$ and $S$. First we observe the following fact.

 \begin{lemma} \label{L3}
 It holds that $\,[\,e^{-it\tau H}\!, P(\tau)\,]
 \stackrel{\scriptsize{s}}{\longrightarrow} 0$ as $\:\tau\to 0$, uniformly
 on each compact $t$-interval in $\R$.
 \end{lemma}
{\em Proof:}
By (\ref{Htau}) with
$\zeta = it\tau$  we have
 $$
  [\,e^{-it\tau H}\!, P(\tau )\,]
   = i\Bigl(P(\tau)t\tau H(it\tau)
         - t\tau H(it\tau)P(\tau) \Bigr),
 $$
and hence for any $v\in \HH$ we can estimate
 $$
  \bigl\|[\,e^{-it\tau H}\!, P(\tau)\,]v \bigr\|
   \leq \|t\tau H(it\tau)v \| + \|t\tau H(it\tau)P(\tau)v \|\,.
 $$
We rewrite (\ref{Htau}) with $\zeta = it\tau$ as
 \begin{equation} \label{decomp-H}
 iH(it\tau) = \frac{I- \cos t\tau H}{t\tau} + i\,
 \frac{\sin t\tau H}{t\tau}
=: B(t\tau) + iA(t\tau)\,,
 \end{equation}
where $B(t\tau)$ and $A(t\tau)$ are obviously bounded self-adjoint
operators on ${\cal H}$, and $B(t\tau)$ is in addition
nonnegative. The definition makes sense if $t\ne0$ but we need not
exclude this case because what we really need is the operator
$t\tau H(it\tau)$. For any $w\in \HH$ we get
 \begin{eqnarray*}
 \|t\tau H(it\tau)w\|^2
 &\!=\!& \|\,[t\tau B(t\tau) + it\tau A(t\tau)]w\|^2\\
 &\!=\!& \|\,[(I-\cos t\tau H) + i\sin t\tau H]w\|^2\\
 &\!=\!& \langle w, [(I-\cos t\tau H)^2 + \sin^2(t\tau H) ]w \rangle \\
 &\!=\!& 4 \|\sin(t\tau H/2) w\|^2 \to 0\,,
 \end{eqnarray*}
uniformly on compact $t$-intervals in $\R $. In this way we have
proved the claim, noting that $P(\tau)
\stackrel{\scriptsize{s}}{\longrightarrow} P$ holds uniformly on
each compact $t$-interval in $\R $ as $\tau \to 0$. \QED

\medskip

Now we employ the following identity,
\begin{eqnarray}
&&\bigl(P(1/n)\,e^{-itH/n}\bigr)^n v -
 \bigl(P(1/n)\,e^{-itH/n}P(1/n)\bigr)^n v \label{diff:ns-s}\\
&&= -\bigl(P(1/n)e^{-itH/n}
          P(1/n)\bigr)^{n-1}[\,e^{-itH/n}, P(1/n)\,]
v\,, \nonumber
\end{eqnarray}
the right-hand side of which converges by Lemma~\ref{L3} to zero
for all $t \not= 0$ and any $v\in\HH$, because
$\bigl(P(1/n)e^{-itH/n}P(1/n)\bigr)^{n-1}$ is a contraction on
$\HH$, and hence also in $L^2_\mathrm{loc}([0,\infty);\, {\cal
H})$. This yields the formula (\ref{product-ns}).

\medskip\noindent
{\it Proof of (\ref{product-ns2}):} In view of the already proven
formula (\ref{product-s}) we have for every $f \in\HH$ and $T>0$
the following chain of relations
\begin{eqnarray*}
 T\|Pf\|^2
     &\geq& \limsup \int_0^T \|(e^{-itH/n}P(1/n))^n f\|^2\, dt \\
     &=& \limsup
     \Big[\int_0^T \|P(1/n)(e^{-itH/n}P(1/n))^n f\|^2 \, dt\\
     &&\qquad\quad
      + \int_0^T \|Q(1/n)(e^{-itH/n}P(1/n))^n f\|^2 \, dt\Big]\\
     &\geq& \limsup
     \int_0^T \|(P(1/n)e^{-itH/n}P(1/n))^n f\|^2 \, dt \\
     &=& \int_0^T \|e^{-itH_P}P f\|^2 \, dt =T\|Pf\|^2,
\end{eqnarray*}
with the $\limsup$ taken along $n\rightarrow \infty$, because $I =
P(1/n) + Q(1/n)$ and $P(\tau)
\stackrel{\scriptsize{s}}{\longrightarrow}P$ as $\tau \rightarrow
0$. It follows that $ \int_0^T \|Q(1/n)(e^{-itH/n}P(1/n))^n f\|^2
\, dt \longrightarrow 0$ as $n \rightarrow \infty$. Thus for any
$v(\cdot) \in L^2_\mathrm{loc}([0,\infty); {\cal H})$ and every
$T>0$ we have, again by (\ref{product-s}),
\begin{eqnarray*}
 &&\int_0^T \langle v(t), (e^{-itH/n}P(1/n))^n f \rangle  \, dt\\
 &=&\int_0^T \langle  P(1/n)v(t),
              (P(1/n)e^{-itH/n}P(1/n))^n f \rangle  \, dt \\
 &&\qquad\quad
     + \int_0^T \langle Q(1/n)v(t),
               Q(1/n)(e^{-itH/n}P(1/n))^n f \rangle  \, dt\\
 &&\longrightarrow
      \int_0^T \langle v(t), e^{-itH_P}Pf \rangle  \, dt
\end{eqnarray*}
as $n\rightarrow \infty$. It means that $\{(e^{-itH/n}P(1/n))^n
f\}$ converges to $e^{-itH_P}Pf$ weakly in
$L^2_\mathrm{loc}([0,\infty); {\cal H})$ together with all the
seminorms, and therefore the convergence is strong in
$L^2_\mathrm{loc}([0,\infty); {\cal H})$. This yields the formula
(\ref{product-ns2}).

\medskip
It remains  to prove Lemma~\ref{L1} on which the above arguments
were based.


\setcounter{equation}{0}

\section{Proof of Lemma~\ref{L1}} \label{s-proof-of-L1}

\noindent To demonstrate (\ref{inv-gen}), we shall use the Vitali
theorem -- see, e.g., \cite{HP} -- for holomorphic functions and
employ arguments analogous to those used in Kato's paper
\cite{Ka2} for the self-adjoint Trotter product formula with the
form sum of a pair of nonnegative self-adjoint operators. We do it
in three steps.

\smallskip\noindent\emph{I.}  In the first step we will show the
following lemma.

 \begin{lemma} \label{L41}
 For a fixed $\zeta = t>0$,
\begin{equation} \label{realconv}
 (I+ S(t,\tau))^{-1} \stackrel{\scriptsize{s}}{\longrightarrow}
 (I + tH_P)^{-1}P \quad \mathrm{as} \quad \tau \to 0\,.
\end{equation}
\end{lemma}
 \smallskip
\noindent {\em Proof:} The argument will be analogous to that in
\cite{Ka2}, and indeed, validity of the result in the particular
case when our projection-valued function is constant is remarked
in \cite[Eq.~(5.2), p.~194]{Ka2}.

For $\zeta = t\tau >0$ we have from (\ref{Htau})
$H(t\tau) = (t\tau)^{-1}[I- e^{-t\tau H}]$, which
is a  bounded,  nonnegative and self-adjoint
operator on ${\cal H}$. It allows us to rewrite
\begin{eqnarray*}
S(t,\tau) &\!=\!& \tau^{-1}[I-  P(\tau)(I-t\tau
H(t\tau))P(\tau)] \\
         &\!=\!& \tau^{-1}Q(\tau) + t P(\tau)H(t\tau)P(\tau)\,,
\end{eqnarray*}
which is in this case also a bounded and nonnegative self-adjoint
operator. To prove (\ref{realconv}) take any $f \in {\cal H}$ and
put $\hat{u}(t,\tau) := (I+ S(t,\tau))^{-1}f$, so that
\begin{equation} \label{fdecomp}
 f =(I + S(t,\tau))\hat{u}(t,\tau) = [I + \tau^{-1}Q(\tau)
 + tP(\tau)H(t\tau)P(\tau)]\hat{u}(t,\tau)\,.
\end{equation}
Then we have
\begin{equation} \label{ufinprod}
 \langle \hat{u}(t,\tau),f \rangle = \|\hat{u}(t,\tau)\|^2
                + \tau^{-1}\|Q(\tau)\hat{u}(t,\tau)\|^2
    + t\|H(t\tau)^{1/2}P(\tau)\hat{u}(t,\tau)\|^2.
\end{equation}
Thus the families $\{\hat{u}(t,\tau)\},\,
\{\tau^{-1/2}Q(\tau)\hat{u}(t,\tau)\}$ and
$\{t^{1/2}H(t\tau)^{1/2} P(\tau) \hat{u}(t,\tau)\}$ are all
bounded by $\|f\|$ for all $t > 0$, uniformly as $\tau\to 0$, and
therefore they are weakly compact in ${\cal H}$. It follows that
for each fixed $t > 0$ there exists a sequence $\{\tau_n (t)\}$
with $\tau_n(t) \to 0$ as $n\to \infty$, in general dependent on
$t$, along which these vectors converge weakly in ${\cal H}$,
\begin{eqnarray} \label{weakconv}
&& \hat{u}(t,\tau) \stackrel{\scriptsize{w}}{\longrightarrow}
\hat{u}(t)\,, \quad
  \tau^{-1/2}Q(\tau)\hat{u}(t,\tau)
   \stackrel{\scriptsize{w}}{\longrightarrow} g_0(t)\,,  \nonumber \\
&& t^{1/2}H(t\tau)^{1/2}P(\tau)\hat{u}(t,\tau)
   \stackrel{\scriptsize{w}}{\longrightarrow} h(t)\,,
\end{eqnarray}
for some vectors $\hat{u}(t), \,  g_0(t)$ and $h(t)$ in
${\cal H}$. Note that the sequence $\{\tau_n(t)\}_{n=1}^{\infty}$
can be chosen the same for all three families.

>From this result we see first that $Q(\tau)\hat{u}(t,\tau)
\stackrel{\scriptsize{s}}{\longrightarrow} 0$ uniformly in $t> 0$
as $\tau\rightarrow 0$, so that we have $Q\hat{u}(t)=0$ or
$\hat{u}(t)=P\hat{u}(t) \in P{\cal H}$. For every $v \in
D[H^{1/2}]$ we have, with the limit taken along $\{\tau_n(t)\}$,
\begin{eqnarray*}
 \langle v, h(t)\rangle
 &\!=\!&  \lim\, \langle v,  t^{1/2}H(t\tau)^{1/2}P(\tau)\hat{u}
(t,\tau) \rangle\\
 &\!=\!& t^{1/2} \,\lim\, \langle H(t\tau)^{1/2}v,P(\tau)\hat{u}
(t,\tau) \rangle
 = t^{1/2} \langle H^{1/2}v, P\hat{u}(t)\rangle\,,
\end{eqnarray*}
because $H(t\tau)^{1/2}v
\stackrel{\scriptsize{s}}{\longrightarrow} H^{1/2}v$ as
$\tau\rightarrow 0$. Hence $\hat{u}(t)=P\hat{u}(t)$ belongs to
$D[H^{1/2}]$ and $h(t) =t^{1/2} H^{1/2}P\hat{u}(t)$ because
$D[H^{1/2}]$ is dense by assumption. Furthermore, multiplying
(\ref{fdecomp}) by $\tau^{1/2}$ and taking the weak limit along
the sequence $\{\tau_n(t)\}$ we get $g_0(t) = 0$.
Similarly, multiplying (\ref{fdecomp}) by $P(\tau)$ we have for
every $v \in D[H^{1/2}P]$
$$
\langle v, P(\tau)f \rangle
 = \langle v, P(\tau) \hat{u}(t,\tau) \rangle
  + \langle t^{1/2}H(t\tau)^{1/2}P(\tau)v,
     t^{1/2}H(t\tau)^{1/2}P(\tau)\hat{u}(t,\tau) \rangle.
$$
Then taking the limit along the sequence $\{\tau_n(t)\}$ we get
$$
  \langle v, Pf \rangle
 = \langle v, P\hat{u}(t) \rangle
  + \langle t^{1/2}H^{1/2}Pv, h(t) \rangle,
$$
because by spectral theorem
\begin{eqnarray*}
 \|H(t\tau)^{1/2}(P(\tau)\!-\! P)v\|
  &\!=\!&  \|H(t\tau)^{1/2}(I+H)^{-1/2}(I+H)^{1/2}(P(\tau)\!-\!P)v\|\\
  &\!\leq\!& \|(I+H)^{1/2}(P(\tau)\!-\! P)v\|\,,
\end{eqnarray*}
which tends to zero since $P(\tau) \stackrel{\scriptsize{s}}
{\rightarrow} P$ as $\tau \to 0$, $\:D[H^{1/2}P(\tau)] \supset
D[H^{1/2}P]$ and $\|H^{1/2}P(\tau)v\| \to \|H^{1/2}Pv\|$ for $v
\in D[H^{1/2}P]$ by assumption\footnote{This part of the proof
shows that the hypotheses of Theorem~\ref{main} can be slightly
weakened, because we need in fact only that
$s-\lim_{\tau\rightarrow 0}\,H(t\tau)^{1/2}P(\tau) v = H^{1/2}P v
$ holds for any $v \in D[H^{1/2}P]$.}. Hence
$H^{1/2}P\hat{u}(t)\in D[H^{1/2}P]$ and
\begin{eqnarray} \label{fu-id}
Pf &=& P\hat{u}(t) + t^{1/2}(H^{1/2}P)^* h(t)
= \hat{u}(t) + t(H^{1/2}P)^* (H^{1/2}P) \hat{u}(t) \nonumber\\
&=& \hat{u}(t) + tH_P \hat{u}(t)\,,
\end{eqnarray}
because $D[H^{1/2}P]$ is supposed to be dense. Applying once again
the standard argument mentioned after Lemma~\ref{L1} to all the
three families we conclude that the weak convergence in
(\ref{weakconv}) takes place independently of a sequence
$\{\tau_n(t)\}$ chosen.

On the other hand, we infer from (\ref{ufinprod}) that
\begin{eqnarray*}
 \langle \hat{u}(t),f \rangle
&\geq& \liminf \|\hat{u}(t,\tau)\|^2
  + \liminf \tau^{-1}\|Q(\tau)\hat{u}(t,\tau)\|^2 \\
 &&\qquad\quad
     + \liminf \|t^{1/2}H(t\tau)^{1/2}P(\tau)\hat{u}(t,\tau)\|^2 \\
&\geq& \|\hat{u}(t)\|^2 + \|g_0(t)\|^2 + \|h(t)\|^2 \\
&=&  \|\hat{u}(t)\|^2 + \|t^{1/2}H^{1/2}P\hat{u}(t)\|^2 \\
&=& \|\hat{u}(t)\|^2 + t\|H_P^{1/2}\hat{u}(t)\|^2
\end{eqnarray*}
with $\liminf$ taken along $\tau \rightarrow 0$. Since by
(\ref{fu-id}) the left-hand side of the above inequality is equal
to
$$\langle \hat{u}(t),f \rangle = \langle \hat{u}(t),Pf \rangle
= \|\hat{u}(t)\|^2 + t\langle \hat{u}(t), H_P\hat{u}(t) \rangle
= \|\hat{u}(t)\|^2 + t\|H_P^{1/2}\hat{u}(t)\|^2,
$$
we see that the norms of these vectors converge to the norms of
their limit vectors. It allows us to conclude that the
$\HH$-valued families in question, $\{\hat{u}(t,\tau)\},\,
\{\tau^{-1/2}Q(\tau)\hat{u}(t,\tau)\}\,$ and
$\{t^{1/2}H(t\tau)^{1/2} P(\tau) \hat{u}(t,\tau)\}$ converge to
$\hat{u}(t)$, $0$ and $t^{1/2}H^{1/2}P\hat{u}(t)$ strongly in
${\cal H} $, respectively, as $\tau\rightarrow 0$. In particular,
we have shown that $Pf = (I+tH_P)\hat{u}(t)$ and $\hat{u}(t,\tau)
\stackrel{\scriptsize{s}}{\longrightarrow} \hat{u}(t) = (I+
tH_P)^{-1}Pf$, or (\ref{realconv}). This proves Lemma~\ref{L41}.
\QED

\medskip\noindent\emph{II.} Next, for a fixed $\tau >0$, the
function $\zeta\mapsto F(\zeta,\tau)$ is holomorphic in the open
right half-plane $\hbox{\rm Re}\, \zeta>0$ and uniformly bounded
in norm by one. This makes it possible to mimick the argument of
Feldman \cite{Fe}, which is reproduced in Chernoff's book
\cite[p.~90]{Ch2}, see also \cite{Fr}, to conclude by means of the
Vitali theorem (see, e.g., \cite[Thm~3.14.1]{HP}) that for
$\hbox{\rm Re}\, \zeta>0$
\begin{equation}
 (I+ S(\zeta, \tau))^{-1}
 \stackrel{\scriptsize{s}}{\longrightarrow} (I+ \zeta H_P)^{-1}P
 \quad \mathrm{as} \quad \tau \rightarrow 0
\end{equation}
holds uniformly on compact subsets of $\hbox{\rm Re}\, \zeta >0$.

At the boundary $\hbox{\rm Re}\, \zeta =0$, or $\zeta = it$ with
$t$ real, $(I+ S(\zeta, \tau))^{-1}$ still converges as $\tau \to
0$ but in a weaker sense only. Using the argument of \cite{Fe}
based on the Poisson kernel, we can check that for each pair of
$f,\, g \in {\cal H}$ and all $\phi \in L^1({\R})$ the following
relation is valid,
\begin{equation} \label{bound_lim}
 s-\lim _{\tau\rightarrow 0}\int_{{\R}} \phi(t)
  \langle g,(I+S(it,\tau))^{-1} f \rangle\, dt
 = \int_{{\R}} \phi(t)
  \langle g,(I+itH_P)^{-1}P f \rangle\, dt\,.
\end{equation}
This says that for each pair of $f,\, g \in {\cal H}$ the family
$\{ \langle g,(I+S(it,\tau))^{-1} f \rangle \}$ of functions of
$t$ in $L^{\infty}({\R})$ converges to $\langle g,(I+itH_P)^{-1}P
f \rangle $ as $\tau \rightarrow 0$ weakly$^*$, or equivalently,
in the weak topology defined by the dual pairing between
$L^{\infty}({\R})$ and $L^{1}({\R})$ -- see, e.g., \cite{Ko}.

\smallskip\noindent\emph{III.} Now we shall show the family of the
bounded operators $\{(I+S(it,\tau))^{-1}\}$ is weakly convergent
in $L^2_\mathrm{loc}([0,\infty);\, {\cal H})$, and in fact,
strongly convergent there too. To do so, we will employ an
argument analogous to that used in the proof of Lemma~\ref{L41} on
the Hilbert space ${\cal H}$, however, this time on the Fr\'echet
space $L^2_\mathrm{loc}([0,\infty); {\cal H})$.

Using the decomposition (\ref{decomp-H}) with $t \not= 0$, we find
(cf. \cite{Ich})
 \begin{eqnarray*}
  S(it,\tau)
 &\!=\!& \tau^{-1} [I- P(\tau)(I-it\tau H(it\tau))P(\tau)] \\
 &\!=\!& \tau^{-1}Q(\tau) + tP(\tau)(B(t\tau) + iA(t\tau))P(\tau)\,.
 \end{eqnarray*}
To prove (\ref{inv-gen}), take any $f \in \HH$ and put
 $
 u(t,\tau) := (I+S(it,\tau))^{-1}f.
 $
Note that this $u(t,\tau)$ represents an element in
$L^2_\mathrm{loc}([0,\infty); {\cal H})$ as well as its unique
representative in $(0,\infty)$, because $u(t,\tau)$ is strongly
continuous at this interval as a function of $t$. Then
 \begin{eqnarray}
 f &\!=\!& (I + S(it,\tau))u(t,\tau) \label{f} \\
 &\!=\!& [I+ \tau^{-1}Q(\tau)  +  tP(\tau)(B(t\tau) +
 iA(t\tau))P(\tau)]u(t,\tau)\,, \nonumber
 \end{eqnarray}
so we  have
 \begin{eqnarray}
  \langle u(t,\tau), f \rangle
          &\!=\!& \langle u(t,\tau),(I+S(it,\tau))u(t,\tau) \rangle
              \nonumber \\
          &\!=\!& \|u(t,\tau)\|^2
            + \tau^{-1}\|Q(\tau) u(t,\tau)\|^2
            + t\|B(t\tau)^{1/2}P(\tau) u(t,\tau)\|^2 \nonumber \\
          && + it\langle P(\tau)u(t,\tau),A(t\tau)P(\tau) u(t,\tau)
\rangle\,.
          \label{uf}
\end{eqnarray}
Observing the real part of (\ref{uf}) we see that for $\tau$ small
enough, each of the ${\cal H}$-valued families $\,\{u(t,\tau)\}$,
$\,\{\tau^{-1/2}Q(\tau) u(t,\tau)\}$ and
$\,\{t^{1/2}B(t\tau)^{1/2}P(\tau) u(t,\tau)\}$ is bounded by
$\|f\|$ for all $t > 0$. Moreover, they are strongly continuous in
$t$ for fixed $\tau >0$, and locally bounded as ${\cal H}$-valued
functions of $t$ in $L^2_\mathrm{loc}([0,\infty); {\cal H}) $,
uniformly as $\tau \rightarrow 0$.

Hence we infer first of all that $Q(\tau) u(t,\tau)
\stackrel{\scriptsize{s}}{\longrightarrow} 0$, uniformly in $t
\in(0,\infty)$, as $\tau \to 0$. Next, since
$L^2_\mathrm{loc}([0,\infty); {\cal H})$ is reflexive
\cite[Chap.~1, Sec.~3.1, pp.~57-62]{GV}, any bounded set in it is
weakly compact \cite[Sec.~23.5, pp.~302-304]{Ko}. Consequently,
there is a sequence $\{\tau_n\}_{n=1}^{\infty}$ with $\tau_n \to
0$ as $n\to \infty$ along which the above families are weakly
convergent in $L^2_\mathrm{loc}([0,\infty); {\cal H})$:
 \begin{eqnarray}
  && u(t,\tau) \stackrel{\scriptsize{w}}{\longrightarrow} u(t)\,,
   \quad  \quad
    \tau^{-1/2} Q(\tau) u(t,\tau)
     \stackrel{\scriptsize{w}}{\longrightarrow} f_0(t)\,, \nonumber \\
  &&  t^{1/2}B(t, \tau)^{1/2}P(\tau) u(t,\tau)
   \stackrel{\scriptsize{w}}{\longrightarrow} z(t)\,,
 \label{weaklim}
 \end{eqnarray}
with some vectors $u(\cdot),\, f_0(\cdot)$ and $z(\cdot) \, \in
L^2_\mathrm{loc}([0,\infty); {\cal H})$. Note that as before the
sequence $\{\tau_n\}_{n=1}^{\infty}$ can be chosen the same for
all three families.

 \smallskip

 \begin{lemma} \label{L5}
These above mentioned vectors have the following properties,
 $$
   u(t) = Pu(t) \in P{\cal H}\,\; \hbox{\rm for a.e.}\; t\,, \quad
   z(\cdot) =0\,, \quad f_0(\cdot) = 0\,.
 $$
\end{lemma}
{\em Proof:} For $B(t\tau)$ and $A(t\tau)$ in (\ref{decomp-H}),
the spectral theorem gives
 \begin{eqnarray}
\|(B(t\tau)^{1/2}v\|^2  &\!=\!&\int_{0-}^{\infty}
    \left|\frac{1-\cos t\tau\lambda}{t\tau\lambda}\right| \|
    E(d\lambda) H^{1/2}v\|^2
    \rightarrow 0\,, \;\; v \in D[H^{1/2}]\,; \nonumber \\
\|(B(t\tau)v\|^2  &\!=\!&\int_{0-}^{\infty}
    \left|\frac{1-\cos t\tau\lambda}{t\tau\lambda}\right|^2 \|
    E(d\lambda) Hv\|^2
    \rightarrow 0\,, \;\; v \in D[H]\, \label{spectral}
 \end{eqnarray}
as $\tau \to 0$ by the Lebesgue dominated-convergence theorem.

Since $Q(\tau) u(t,\tau) \stackrel{\scriptsize{s}}
{\longrightarrow} 0$ uniformly in $t \in(0,\infty)$ and $Q(\tau)
\stackrel{\scriptsize{s}}{\longrightarrow} Q$ as $\tau \to 0$, we
have $Qu(t)=0$, or in other words $u(t)=Pu(t) \in P{\cal H}$ for
a.e. $t$. Moreover, by (\ref{spectral}) we infer that
 \begin{eqnarray*}
  \int_0^{\infty} \langle \phi (t) v,z(t) \rangle \, dt
 &\!=\!& \lim \int_0^{\infty}
      \langle \phi (t) v, t^{1/2}B(t\tau)^{1/2}P(\tau) u(t,\tau)
\rangle \, dt\\
 &\!=\!& \lim \int_0^{\infty} \bar{\phi} (t)
  \langle  t^{1/2}B(t\tau)^{1/2}v,P(\tau) u(t,\tau) \rangle \, dt\\
 &\!=\!& \int_0^{\infty} \bar{\phi} (t) \langle 0,Pu(t) \rangle \, dt = 0
 \end{eqnarray*}
holds for every $\phi \in C_0^{\infty}([0,\infty))$ and $v \in
D[H^{1/2}]$, hence $z(t)=0$  a.e. because $D[H^{1/2}]$ is dense in
$\HH$, so that $z(\cdot)$ is the zero element of
$L_\mathrm{loc}^2([0,\infty); {\cal H})$. Finally, the relation
$f_0(\cdot)=0$ follows from (\ref{f}) which implies
$\tau^{1/2}Q(\tau)f = \tau^{1/2}(1+\tau^{-1})Q(\tau)u(t,\tau)$,
yielding the result; this concludes the proof. \QED

\bigskip

Our next aim is to show that the weak limits in (\ref{weaklim}) do
not depend upon a sequence chosen. The  $u(\cdot,\tau_n) =
(I+S(it,\tau_n))^{-1}f$ converge to $u = u(\cdot)$ weakly in
$L^2_\mathrm{loc}([0,\infty); {\cal H})$ as $n\to \infty$. It
obviously implies that for all $\phi  \in
C_0^{\infty}([0,\infty))$ and for every $g \in {\cal H}$ we have
$$
 \int_0^{\infty} \phi(t) \langle g,(I+ S(it,\tau))^{-1}f \rangle
 \, dt\; {\longrightarrow} \int_0^{\infty}\phi(t) \langle g, u(t) \rangle
 \, dt\,,
$$
again along the sequence $\{\tau_{n}\}$. It follows from
(\ref{bound_lim}) that
\begin{equation} \label{repr-of-u(t)}
u(t) = (I+itH_P)^{-1}Pf, \quad \hbox{\rm for}\,\,
\hbox{\rm a.e.}\,\, t \,\, \hbox{\rm in}\,\, [0,\infty),
\end{equation}
because the set of all such $\bar{\phi}(\cdot) g$ is total in
$L^2_\mathrm{loc}([0,\infty); {\cal H})$. This shows that for
every $f \in {\cal H}$, $(I+S(it,\tau_n))^{-1}f$ converges to
$(I+itH_P)^{-1}Pf$ weakly in $L_\mathrm{loc}^2([0,\infty); {\cal
H})$ as $n\rightarrow \infty$. Together with the fact that
$z(\cdot)=0,\, f_0(\cdot)=0$ in view of Lemma~\ref{L5}, this
yields the desired property, namely that the weak limits of
(\ref{weaklim}) are independent of the particular subsequence
$\{\tau_n\}$ chosen. The standard argument sketched below
Lemma~\ref{L1} shows that (\ref{weaklim}) holds as $\tau
\rightarrow 0$ without any restriction on subsequences.

\bigskip
Finally, we are going to check the strong convergence
$u(\cdot,\tau) \stackrel{\scriptsize{s}}{\longrightarrow}
u(\cdot)$ in $L_\mathrm{loc}^2([0,\infty); {\cal H})$ as
$\tau\rightarrow 0$. In fact, we will prove two other limiting
relations at the same time.

 \begin{lemma} \label{L61}
 In the topology of $L_\mathrm{loc}^2([0,\infty); {\cal H})$,
 the family $\{u(\cdot,\tau)\}$ converges to the vector
 $u = u(\cdot)$ as $\tau\rightarrow 0$, and moreover,
 \begin{eqnarray*}
 && \tau^{-1/2}Q(\tau) u(t,\tau)
   {\:\longrightarrow\:} f_0(t) = 0\,, \\
 &&  t^{1/2}B(t, \tau)^{1/2}P(\tau) u(t,\tau)
     {\:\longrightarrow\:} z(t) = 0\,.
 \end{eqnarray*}
 \end{lemma}
{\em Proof:} In the above reasoning we have already checked the
weak convergence in (\ref{weaklim}) as $\tau\rightarrow 0$.
Integrating the real part of (\ref{uf}) in $t$ over the interval
$[0,T]$ for any fixed $T>0$ and taking $\liminf$ as
$\tau\rightarrow 0$, we get by Lemma~\ref{L5}
\begin{eqnarray*}
 \hbox{\rm Re} \int_0^T \langle u(t), f \rangle \, dt
   &\geq& \liminf \int_0^T \|u(t,\tau)\|^2 \, dt \\
   &&\qquad
      +  \liminf \int_0^T \tau^{-1}\|Q(\tau) u(t,\tau)\|^2 \, dt\\
     &&\qquad\quad
      + \liminf \int_0^T  \|t^{1/2}B(t\tau)^{1/2}P(\tau) u(t,\tau)
\|^2 \, dt \\
    &\geq& \int_0^T \|u(t)\|^2 \, dt + \int_0^T \|f_0(t)\|^2 \, dt
          + \int_0^T \|z(t)\|^2 \, dt   \\
    &\geq& \int_0^T \|u(t)\|^2 \, dt\,.
\end{eqnarray*}
On the other hand, the left-hand side of the above inequality is
by (\ref{repr-of-u(t)}) equal to
$$
 \hbox{\rm Re} \int_0^T \langle u(t), f \rangle \, dt
 = \hbox{\rm Re} \int_0^T \langle u(t), (I+ itH_P)u(t) \rangle \, dt
 = \int_0^T \|u(t)\|^2\, dt.
$$
Hence we conclude that all the Fr\'echet-space semi-norms of the
vectors $u(t,\tau)$, $\tau^{-1/2}Q(\tau) u(t,\tau)$ and
$t^{1/2}B(t,\tau)^{1/2}P(\tau)u(t,\tau)$ converge to the
semi-norms of the weak-limit vectors $u(t),\, 0$ and $0$,
respectively, as $\tau\rightarrow 0$. Thus the convergence is
strong with respect to each semi-norm, and since their family
induces the topology in $L_\mathrm{loc}^2([0,\infty); {\cal H})$
the lemma is proved. \QED

\medskip

This completes the proof of Lemma~\ref{L1}, and by that the
verification of our main result, Theorem~\ref{main}.


\setcounter{equation}{0}

\section{The finite-dimensional case} \label{finite}

In this section, we will prove Theorem~\ref{fin} in which we
assume that $P$ and $P(t)$ are finite-dimensional orthogonal
projections. Since the closed operator $H^{1/2}P$ is supposed to
be densely defined, the domain $D[H^{1/2}P]$ of $H^{1/2}P$ becomes
the whole space ${\cal H}$, for the restriction $H^{1/2}P|_{P{\cal
H}}$ of the operator $H^{1/2}P$ to the finite-dimensional subspace
$P{\cal H}$ is densely defined, so its domain must coincide with
$P{\cal H}$, and it acts as zero on $Q\HH$. The same is valid for
$H^{1/2}P(t)$ when $P(t)$ is of a finite dimension. As a result,
$H^{1/2}P$ and $H_P = (H^{1/2}P)^* (H^{1/2}P)$ as well as
$H^{1/2}P(t)$ are bounded operators on ${\cal H}$ by the
closed-graph theorem. By the assumptions common with
Theorem~\ref{main}, for each fixed $f \in D[H^{1/2}P] = {\cal H}$
the family $\{H^{1/2}P(t)f\}$ converges to $H^{1/2}Pf$ as $t
\rightarrow 0$, and hence is uniformly bounded with respect to $t$
near to zero, say, for $-1 \leq t \leq 1$. Then by the uniform
boundedness principle we can conclude that $\sup_{|t| \leq 1}
\|H^{1/2}P(t)\| < \infty$.

To prove the assertions (i) and (ii) of Theorem~\ref{fin}
simultaneously, take a fixed $a \in \R$ and consider instead of
$F(\zeta,\tau),\, S(\zeta,\tau)$ defined by (\ref{red-evol}) and
(\ref{red-evol-der}), respectively, the following operators
$$
 F_a(\zeta,\tau) := P(a\tau)\exp(-\zeta\tau H)P(a\tau),
 \quad S_a(\zeta,\tau) := \tau^{-1}[I-F_a(\zeta,\tau)].
$$
In fact, we shall employ $F_a(it,\tau),\, S_a(it,\tau)$ instead of
$F(it,\tau),\, S(it,\tau)$ in the proof of Lemma~\ref{L2} and
Lemma~\ref{L1}. Similarly $u(t,\tau)$ used above will be replaced
by $u_a(t,\tau) = (I+S_a(it,\tau))^{-1}f$ corresponding to a given
$f\in\HH$.

\begin{lemma} \label{L7}
For any $t, t' \geq 0$ and $0< \tau \leq 1$ we have
$$
  \|u_a(t,\tau) - u_a(t',\tau) \| \leq C(a)|t-t'|\, \|f\|
$$
with a positive $C(a)$ independent of $t, t'$, which is uniformly
bounded as a function of $a$ on each compact interval of $\,\R$.
\end{lemma}
\emph{Proof:} By the resolvent equation we have
\begin{eqnarray*}
 \lefteqn{u_a(t,\tau) - u_a(t',\tau)} \\
 &&= (I+S_a(it,\tau))^{-1}f - (I+S_a(it',\tau))^{-1}f\\
 &&= (I+S_a(it,\tau))^{-1}
      P(a\tau) \tau^{-1}[e^{-it\tau H} - e^{-it'\tau H}]P(a\tau)
     (I+S_a(it',\tau))^{-1}f\\
 &&= (I+S_a(it,\tau))^{-1}
      P(a\tau)\, \frac{1}{\tau}\, \int_{t'}^t \frac{d}{ds}
      \,e^{-is\tau H} ds\;
       P(a\tau)(I+S_a(it',\tau))^{-1}f\\
 &&= -i (I+S_a(it,\tau))^{-1}
      P(a\tau) \int_{t'}^t H e^{-is\tau H} ds\;
       P(a\tau)(I+S_a(it',\tau))^{-1}f\\
 &&= -i (I+S_a(it,\tau))^{-1}
      (H^{1/2}P(a\tau))^* \\ && \phantom{AA}
      \times \int_{t'}^t e^{-is\tau H} ds
      \; (H^{1/2}P(a\tau))(I+S_a(it',\tau))^{-1}f\,.
\end{eqnarray*}
At the beginning of this section we have argued that the operators
$H^{1/2}P(a\tau)$ are uniformly bounded on ${\cal H}$ for $0 <
\tau \leq 1$. It follows that
$$
 \|u_a(t,\tau) - u_a(t',\tau)\| \leq C(a) |t-t'|\, \|f\|
$$
with $C(a) := \sup_{|a\tau| \leq 1} \|H^{1/2}P(a\tau)\|^2$. By the
argument preceding the lemma the function $C(\cdot)$ is uniformly
bounded on each compact $a$-interval in $\R$; this yields the
claim. \QED

\medskip

\noindent \emph{Proof of Theorem~\ref{fin}:} It follows from the
lemma that the vector family $\{u_a(t,\tau)\}$, continuous in ${\cal
H}$, is uniformly bounded and equicontinuous. Hence we may infer
by the Ascoli--Arzel\`a theorem that the sequence $\{\tau_n\}$
used in part III of the proof of Lemma~\ref{L1} can chosen to have
an additional property, namely that the sequence $\{u_a(t,\tau_n)\}$
converges strongly to $u(t)$ also pointwise, uniformly on
$[0,\infty)$. Then the limit $u(t)$ becomes strongly continuous in
$t \geq 0$, and coincides with $(I+itH_P)^{-1}f$ for all $t \geq
0$. Thus we have instead of Lemma~\ref{L1} the following claim:
\begin{equation} \label{5-resovconv0}
  (I+S_a(it,\tau))^{-1} \longrightarrow (I+itH_P)^{-1}P
\end{equation}
as $\tau \rightarrow 0$, strongly on ${\cal H}$ and uniformly on
each compact interval of the variable $t$ in $[0,\infty)$.

Next we will modify the reasoning of Sec.~\ref{s-proof} based on
\cite[Theorem 1.1]{Ch2} with the aim to show the symmetric product
case,
\begin{equation} \label{f-dimTrotter}
 [P(at/n)\exp(-itH/n)P(at/n)]^n  \stackrel{\scriptsize{s}}
 {\longrightarrow}\exp(-itH_P)P,  \quad n \rightarrow \infty.
\end{equation}
Let $f \in\HH$. The resolvent convergence (\ref{5-resovconv0})
with $t=1$ implies the convergence of the corresponding
semigroups, so we have
\begin{equation} \label{5-semigroupconv1}
 e^{-\theta S_a(i, \tau)}f \stackrel{\scriptsize{s}}{\longrightarrow}
 e^{-i\theta H_P}f
\end{equation}
in $\HH$ as $\tau\to 0$, uniformly on each compact interval of the
variable $\theta \geq 0$. Using this equivalence once more we get
for any $\lambda \geq 0$ the relation
$$
 (I+ \lambda S_a(i, \tau))^{-1}f
 \stackrel{\scriptsize{s}}{\longrightarrow}
 (I+i\lambda H_P)^{-1}Pf\,, \quad \tau \rightarrow 0.
$$
In particular, taking $\tau = \theta/n$ and using the diagonal
trick, we infer that
\begin{equation} \label{5-resolvconv2}
 (I+ \frac{\theta}{\sqrt{n}} S_a(i, \theta/n))^{-1}f \stackrel{\scriptsize{s}}
 {\longrightarrow}Pf\,,  \quad n \rightarrow \infty\,,
\end{equation}
holds uniformly on each compact $\theta$-interval in $[0,\infty)$.
Then the mentioned lemma from \cite{Ch1} yields
$$
\left\|F_a(i, \theta/n)^{n} g - e^{-n(I-F_a(i,\theta/n))}g\right\|
 \leq \sqrt{n}\left\|(I-F_a(i, \theta/n))g\right\|\,.
$$
Choosing again $g = \left(I+ \frac{\theta}{\sqrt{n}} S_a(i,
\theta/n)\right)^{-1}\!f\:$ we find that
\begin{eqnarray*}
&&\left\|\left[ F_a(i, \theta/n)^{n} - e^{-\theta S_a(i,\theta/n)}\right]
       \left(1+ \frac{\theta}{\sqrt{n}} S_a(i, \theta/n)\right)^{-1}f
       \right\| \\
&& \quad \leq
 \left\|\left(I+ \frac{\theta}{\sqrt{n}} S_a(i, \theta/n)\right)^{-1}f
 - f\right\|\,,
\end{eqnarray*}
where the right-hand side tends to zero as $n \rightarrow \infty$
by (\ref{5-resolvconv2}). Using the last named convergence once
more we get
\begin{equation} \label{5-semigrconv2}
\left\|F_a(i, \theta/n)^{n} f - e^{-\theta S_a(i,\theta/n)}f \right\|
 \longrightarrow  0\,
\end{equation}
uniformly on each compact $\theta$-interval in $[0,\infty)$.
Choosing now $\theta = t$ we see that the validity of
(\ref{f-dimTrotter}) on $P{\cal H}$ follows immediately from
(\ref{5-semigroupconv1}) and (\ref{5-semigrconv2}). Consequently,
on the subspace $P\HH$ the assertion (i) is obtained by taking
$a=1/t$ for any $t$ belonging to a compact interval in $\R
\setminus \{0\}$ and (ii) by choosing simply $a=1$.

The case $f \in Q{\cal H}$ can be treated as in the proof of
Lemma~\ref{L2}; together this yields the relation
(\ref{f-dimTrotter}) on ${\cal H}$, i.e. the symmetric product
case. The non-symmetric product cases can also be checked with the
help of Lemma~\ref{L3} -- cf.~part (b) of the proof of
Theorem~\ref{main} in Section~3. This concludes the proof
of Theorem~\ref{fin}. \QED


\setcounter{equation}{0}

\section{An example} \label{s-example}

As we have said, our investigation was motivated by the result by
Facchi et al. \cite{FP} mentioned in the introduction. Let us thus
look how the result looks in this case. To see this, consider an
open domain $\Omega\subset\R^d$ with a smooth boundary, and denote
by $P$ the orthogonal projection on $L^2(\R^d)$ defined as the
multiplication operator by the indicator function $\chi_\Omega$ of
the set $\Omega$. Consider further the free quantum Hamiltonian $H
:= -\Delta$, i.e. the Laplacian in $\R^d$ which is a nonnegative
self-adjoint operator in $L^2(\R^d)$, and the Dirichlet Laplacian
$-\Delta_\Omega$ in $L^2(\Omega)$ defined in the usual way
\cite[Sec.~XIII.15]{RS} as the Friedrichs extension of the
appropriate quadratic form.

We consider the Zeno dynamics in the subspace $L^2(\Omega)$
corresponding to a permanent reduction of the wavefunction to the
region $\Omega$, which may be identified with the volume of a
detector. In the sense of the $L_\mathrm{loc}^2(\R; L^2({\R}^d))$
topology, which is physically plausible as explained in
Remark~\ref{phys-conv}, we then claim that the generator of the
dynamics in $L^2(\Omega)$ is just the appropriate Dirichlet
Laplacian,
 \begin{equation}
     (Pe^{-it(-\Delta/n)}P)^n
     \to e^{-it(-\Delta_\Omega)}P
 \end{equation}
as $n\to\infty$, or in other words:

 \begin{proposition}
 The self-adjoint operator
 \begin{equation}
 -\Delta_P = ((-\Delta)^{1/2}P)^*((-\Delta)^{1/2}P)
 \end{equation}
is densely defined in $L^2(\R^d)$ and its restriction to the
subspace $L^2(\Omega)$ is nothing but the Dirichlet Laplacian
$-\Delta_\Omega$ of the region $\Omega$, with the domain
$D[-\Delta_\Omega] = W_0^1(\Omega) \cap W^2(\Omega)$.
  \end{proposition}
{\em Proof:} Let $u \in D[-\Delta_P]$, so that $u$ and
$-\Delta_Pu$ belong to $L^2(\R^d)$. We have
 $$ \langle -\Delta_P u, \varphi\rangle
 = \langle u, -\Delta \varphi \rangle
 =  \langle -\Delta u, \varphi\rangle,
 $$
for any $\varphi \in C_0^{\infty}(\Omega)$ because $\varphi$ has a
compact support in $\Omega$. Thus $-\Delta_Pu = -\Delta u$ holds
in $\Omega$ in the sense of distributions, which means that
$\Delta u|_\Omega \in L^2(\Omega)$. On the other hand, since
$(-\Delta)^{1/2}P u \in L^2(\R^d)$, we have $\chi_{\Omega} u \in
W^1(\R^d)$. Since we have
 $$
 \nabla(\chi_\Omega u) = \nabla((\chi_\Omega)^2 u)
  =  (\nabla \chi_\Omega) \chi_\Omega u(x)
    + \chi_\Omega \nabla (\chi_\Omega u),
 $$
in order to belong to $L^2(\R^d)$ the function $\nabla(\chi_\Omega
{u})$ must not contain the $\delta$-type singular term, which
requires $u(\cdot) = 0$ on the boundary of $\Omega$.
 This combined with the fact that $u|_{\Omega}, \, \Delta u|_{\Omega}
\in L^2(\Omega)$ -- see, e.g., \cite[Thm~5.4]{LM} -- implies that
$u|_{\Omega}$ belongs to $W^2(\Omega)$ and $W_0^1(\Omega)$.

Thus we have shown that $u|_\Omega \in D[-\Delta_\Omega]$ and
$(-\Delta_Pu)|_{\Omega} = -\Delta_\Omega (u|_{\Omega})$
or $-\Delta_\Omega\supset -\Delta_P|_{L^2(\Omega)}$, but both
operators are self-adjoint, so they coincide. \QED

\medskip

In this sense therefore our result given in Theorem~\ref{main}
provides one possible abstract version of the result by Facchi et
al. \cite{FP}.







\subsection*{Acknowledgments}

P.E. and T.I. are respectively grateful for the hospitality
extended to them at Kanazawa University and at the Nuclear Physics
Institute, AS CR, where parts of this work were done. The research
has been partially supported by ASCR and Czech Ministry of
Education under the contracts K1010104 and ME482, and by the
Grant-in-Aid for Scientific Research (B) No.~13440044 and
No.~16340038, Japan Society for the Promotion of Science.


\bigskip

\begin{thebibliography}{99}
 \bibitem[AS]{AS}
 M.S.~Abramowitz and I.A.~Stegun, eds.: {\em Handbook of Mathematical
 Functions}, Dover, New York 1965.
 \vspace{-1.8ex}
 \bibitem[BN]{BN}
 J.~Beskow and J.~Nilsson: The concept of wave function and the
 irreducible representations of the Poincar\'{e} group, II.~Unstable
 systems and the exponential decay law, {\em Arkiv Fys.} {\bf 34}
 (1967), 561-569.
 \vspace{-1.8ex}
 \bibitem[Ch1]{Ch1}
 P. R.~Chernoff: Note on product formulas for operator semigroups,
 {\em J. Funct. Anal.} {\bf 2} (1968), 238--242.
 \vspace{-1.8ex}
 \bibitem[Ch2]{Ch2}
 P. R.~Chernoff: {\em Product Formulas, Nonlinear Semigroups, and
 Addition of Unbounded Operators}, Mem. Amer. Math. Soc. {\bf 140};
 Providence, R.I. 1974.
 \vspace{-1.8ex}
 \bibitem[Ex]{Ex}
 P.~Exner: {\em Open Quantum Systems and Feynman Integrals}, D.~Reidel
 Publ. Co., Dordrecht 1985
 \vspace{-1.8ex}
 \bibitem[EI]{EI}
 P.~Exner and T.~Ichinose: Product formula for quantum Zeno dynamics,
 to appear in \emph{Proceedings of the XIV International
 Congress of Mathematical Physics (M$\,\cap\,\Phi$), Lisbon,
 July 28--Aug 2, 2003.}
 \vspace{-1.8ex}
 \bibitem[FPS]{FP}
 P.~Facchi, S.~Pascazio, A.~Scardicchio, and L.S.~Schulman:
 Zeno dyna\-mics yields ordinary constraints, {\em Phys. Rev.}
 {\bf A 65} (2002), 012108.
 \vspace{-1.8ex}
\bibitem[Fe]{Fe}
 J.~Feldman: On the Schr\"odinger and heat equations for nonnegative
 potentials, {\em Trans. Amer. Math. Soc.} {\bf 108} (1963),
 251--264.
 \vspace{-1.8ex}
 \bibitem[Fr]{Fr}
 C. Friedman:  Semigroup product formulas, compressions, and
 continual observations in quantum mechanics, {\em Indiana Math. J.}
 {\bf 21} (1971/72), 1001--1011.
 \vspace{-1.8ex}
 \bibitem[GV]{GV}
 O.M.~Gel'fand and N.Y.~Vilenkin: {\em Generalized Functions,
 IV.~Applications of Harmonic Analysis}, Academic Press, New York
1965.
 \vspace{-1.8ex}
 \bibitem[HP]{HP}
 E. Hille and R. S. Phillips:
 {\em Functional Analysis and Semi-groups}, Amer. Math. Soc.
  Colloquium Publ. No. 31, rev. ed., Providence, R. I. 1957.
 \vspace{-1.8ex}
 \bibitem[Ich]{Ich}
 T.~Ichinose: A product formula and its application to the
 Schr\"odinger equation, {\em Publ. RIMS Kyoto Univ.} {\bf 16} (1980),
 585--600.
 \vspace{-1.8ex}
 \bibitem[Ka1]{Ka1}
 T.~Kato: {\em Perturbation Theory for Linear Operators}, Springer,
 Berlin-Heidelberg-New York 1966.
 \vspace{-1.8ex}
 \bibitem[Ka2]{Ka2}
 T.~Kato: Trotter's product formula for an arbitrary pair of
 self-adjoint contraction semigroups, in {\em Topics in Functional
 Analysis} (I. Gohberg and M. Kac, eds.),
 Academic Press, New York 1978;  pp.185--195.
 \vspace{-1.8ex}
 \bibitem[K\"o]{Ko}
 G.~K\"othe: {\em Topological Vector Spaces I}, Springer,
 Berlin-Heidelberg-New York 1969.
 \vspace{-1.8ex}
 \bibitem[LM]{LM}
 J.~L.~Lions and E.~Magenes: {\em Non-Homogeneous Boundary Value
 Problems and Applications I}, Springer,$\:$Berlin-Heidelberg-New York 1972.
 \vspace{-1.8ex}
 \bibitem[MaS]{MaS}
 M.~Matolcsi and R.~Shvidkoy: Trotter's product formula for
 projections, {\em Arch. der Math.} {\bf 81} (2003), 309--317.
 \vspace{-1.8ex}
 \bibitem[MS]{MS}
 B.~Misra and E.C.G.~Sudarshan: The Zeno's paradox in
 quantum theory, {\em J. Math. Phys.} {\bf 18} (1977), 756--763.
 \vspace{-1.8ex}
 \bibitem[RS]{RS} M.~Reed and B.~Simon: {\em Methods of Modern
 Mathematical Physics, IV. Analysis of Operators,}
 Academic Press, New York 1978.
 \vspace{-1.8ex}

 \end{thebibliography}
\end{document}